\documentclass[twocolumn,showpacs,preprintnumbers,amsmath,amssymb,superscriptaddress,aps,prc]{revtex4-1}
\usepackage{graphicx}
\usepackage{xcolor}
\usepackage{subfigure}
\def\scr{\rm\scriptscriptstyle }

\begin{document}
\title{Assessing the adequacy of the bare optical potential in near-barrier fusion calculation}
\author{L. F. Canto}
\affiliation{Instituto de F\'{\i}sica, Universidade Federal do Rio de Janeiro, CP 68528,
Rio de Janeiro, Brazil}
\author{P. R. S. Gomes}
\affiliation{Instituto de F\'{\i}sica, Universidade Federal Fluminense, Av. Litoranea
s/n, Gragoat\'{a}, Niter\'{o}i, R.J., 24210-340, Brazil}
\author{J. Lubian}
\affiliation{Instituto de F\'{\i}sica, Universidade Federal Fluminense, Av. Litoranea
s/n, Gragoat\'{a}, Niter\'{o}i, R.J., 24210-340, Brazil}
\author{M. S. Hussein}
\affiliation{Instituto de Estudos Avan\c{c}ados, Universidade de S\~{a}o Paulo C. P.
72012, 05508-970 S\~{a}o Paulo-SP, Brazil, and Instituto de F\'{\i}sica,
Universidade de S\~{a}o Paulo, C. P. 66318, 05314-970 S\~{a}o Paulo,-SP,
Brazil}
\author{P. Lotti}
\affiliation{INFN, Sezione di 
Padova\\
Via F. Marzolo 8, I-35131, Padova, Italy}

\date{\today}

\begin{abstract}
We critically examine the differences among the different bare nuclear interactions used in near-barrier heavy ion fusion analysis and 
Coupled-Channels calculations, and discuss the possibility of extracting the barrier parameters of the bare potential from above-barrier 
data. We show that the choice of the bare potential may be critical for the analysis of the fusion cross sections. We show also that the 
barrier parameters taken from above barrier data may be very wrong.
\end{abstract}

\maketitle
\section{Introduction}
The accepted view of heavy-ion fusion reactions analyses, developed over the last several decades, is a combination of one-dimensional 
barrier penetration model, described by an appropriate ``bare" optical potential, augmented by couplings to important channels whose 
coupling with the entrance channel is required to account for the data \cite{CGD06}. The choice of the bare potential is important as it sets 
the stage for a critical assessment of the importance of including the coupling to non-elastic channels in the calculation of the fusion cross
 section and confronting this with the data. This is most  pronounced at energies in the vicinity of the height of the Coulomb barrier, where
 tunneling is dominant. At higher energies the fusion cross section follows a simple geometrical form, with a linear dependence on the 
 inverse of the collision energy in the centre of mass frame, $E$. The choice of the bare potential is also important as it dictates the angular
 and the linear coefficients of this 1/E dependence (see Sec. III).

\medskip

Coupled channels effects on the fusion cross section at high energies manifest themselves in the form of an average energy loss. 
When the energy is increased further, many channels become open in the region of the so-called deep inelastic reactions, and the fusion 
cross section starts decreasing after reaching a maximum. In this same region competition with breakup reactions become important. This 
is the accepted scenario of the fusion of tightly bound nuclei which was under experimental scrutiny over many decades now.

At near or below barrier energies the fusion of these tightly bound nuclei is dictated by tunneling and the effect of the coupling to collective degrees of freedom in the participating nuclei is most important, giving rise to great enhancement of the sub-barrier fusion when compared to that calculated with just the one-channel, bare potential. 

The low energy fusion of weakly bound projectiles is more challenging as the effect of the coupling to the breakup channel even at near 
barrier energies is very important and yet quite difficult to calculate precisely with available coupled channels codes such as the Continuum
Discretized Coupled Channels (CDCC) one~\cite{HVD00,DiT02,DTB03}. In such a situation one relies heavily on a comparison with the data 
of a very precise calculation of the fusion cross section which takes into account all the important bound non-elastic channels using an 
appropriate bare optical potential. The difference with the data is then attributed to the effect of the coupling to the breakup channels. It then
 becomes very clear how important the choice of the bare potential is. A poor choice can lead to  conflicting conclusions especially in the case
 of halo nuclei such as $^{6}$He.

\section{The bare potential}

There are several bare potentials which are used in the near-barrier fusion analysis. The most frequently used ones are briefly discussed
below.

\medskip

\begin{enumerate}

\item{Phenomenological}

Several analyses have been performed using the Wood-Saxon form for the bare potential. For the record we give this form,
\begin{equation}
V^{\scr WS}(r) = \frac{V_0}{1 + \exp\left[ \left( r - R_0 \right)/a \right] }.
\end{equation}
Usually, a similar form is taken for the imaginary part, plus a surface term which takes into account absorption at the boundary of 
the nuclear system. \\

\item{Microscopic-inspired double folding bare potential}

The double folding form of the bare potential is inspired by a low-energy version of multiple scattering theory, with due respect given to 
medium modifications. Its general form is,

\begin{equation}
V^{\scr DF}(r) = \int d{ \bf r}^{\prime}\,d {\bf r}^{\prime\prime}\ \rho_{\scr P}({ \bf r}^{\prime})\, v({\bf r} - {\bf r}^{\prime} + 
{\bf r}^{\prime\prime})\, \rho_{\scr T}(\bf{r}^{\prime\prime}),
\end{equation}

where $\rho_{\scr P}$ and $\rho_{\scr T}$ are the densities of the projectile and the 
target, respectively. The interaction $v$ is a configuration version of the Bruckner 
G-matrix~\cite{KuB66}. The well known choice is the M3Y interaction~\cite{BBM77}. A version of the double folding potential which takes 
into account the effect of Pauli non-locality in $v$ is the S\~ao Paulo (SP) potential~\cite{CPH97,CCG02}, which has been used extensively in recent years in the analysis of heavy-ion collisions. It relies on the single folding potential of nucleon-nucleus interaction, with appropriate use of the non-locality ala Perey-Perey and Buck and Perey~\cite{PeB62}. The single folding potential is then folded into the density of the projectile. This gives the following form of the S\~ao Paulo potential,
\begin{equation}
V^{\scr SP}(r) = V^{\scr DF}(r)\  e^{- {\rm v}^{2}(r)/ 4c^2 },
\end{equation}
where ${\rm v}(r)$ is the local relative velocity. The S\~ao Paulo potential at near-barrier energies reduces to the double folding one, as the non-locality factor $\exp\left[ - {\rm v}^{2}(r)/4c^2\right] $ becomes insignificant in this regime. The merit of the S\~ao Paulo bare potential resides in the fact that the double folding integral is evaluated exactly with any kind of density.

\item{The Aky\"uz-Winther potential}

This bare potential is an approximation of the double folding interaction which supplies an analytical form. It invloves the use of a Fermi-type functions for the densities and the M3Y interaction. The double folding integral is then evaluated. The potential for a broad range of systems are then approximated by the WS functions,

\begin{equation}
V^{\scr AW}(r) = C \left[\frac{R_{\scr P}R_{\scr T}}{R}\right]\frac{1}{1 + \exp{[\kappa (r - R)]}},
\end{equation}
where 
\[ 
C=-0.65.4\, {\rm MeV/fm}, \qquad R = R_{\scr P} + R_{\scr T}, 
\]
and
\begin{eqnarray}
R_i &=& \left[1.2 A_{i}^{1/3} - 0.35\right]  {\rm fm}, \ \ (i={\rm P}, {\rm T}),\\
\kappa &=& \left[1.16 + 0.56\left(\frac{1}{A_{P}^{1/3}} + \frac{1}{A_{T}^{1/3}}\right)\right]  {\rm fm}^{-1}.
\end{eqnarray}

The AW bare potential is a very convenient interaction as it has an analytic form, with parameters given for any nuclear system and thus 
easy to use in reaction calculation.

\item{The Proximity potential} 

The Proximity (Prox) potential~\cite{BRS77} is based on the approximation that the diffuseness of the nuclear surface is 
much smaller than the radius. This is the case for leptodermic systems. As such one can approximate the interaction 
between the projectile and the target at small separation by that of two slabs of nuclear matter in the sense,
\begin{equation}
V^{\scr prox}(r) = 2\pi\int_{0}^{\infty}dr_{\bot}\  r_{\bot} \, E(r_{\bot}),
\end{equation}
where $E(r_{\bot})$ is the interaction per unit area of two semi-infinite surfaces at the distance $r_{\bot}$.

The interaction above can be put in the following form,
\begin{equation}
V^{\scr prox}(r) = 4\pi\gamma\, b\,  {\bar R}\  \Phi(\zeta) ,
\label{Phi}
\end{equation}
with, $b = 1$ fm, and ${\bar R} = 1/\left( R_{\scr P}^{-1} + R_{\scr T}^{-1} \right)$, and
\begin{equation}
R_i = \left[ 1.28\, A_i^{1/3} - 0.76 + 0.8 A_i^{-1/3}\right]  {\rm fm}.
\end{equation}
The universal function $\Phi(\zeta)$ is given by
\begin{equation}
\Phi(\zeta) = -0.5\, \left( \zeta - 2.54 \right)^2 - 0.0853 \, \left( \zeta - 2.54 \right)^3, 
\end{equation}
for $\zeta \leq 1.2511$, while for $\zeta \geq 1.2511$ it is
\begin{equation}
\Phi(\zeta) = -3.437 e^{-\zeta/0.75}.
\end{equation}

The Proximity potential is convenient and easy to use. However, it is most appropriate to heavier systems in line 
with the leptodermic approximation. Nevertheless we will use the Proximity potential in the comparison with the SP 
and the AW ones, in the next section. The comparison among  the three bare potentials, the Proximity, the SP, and 
the AW, will serve as a benchmark for our discussion of the fusion cross section of loosely bound nuclei.
\end{enumerate}

\bigskip

\section{Coulomb barriers  and fusion cross sections}

\begin{figure}[th]
\centering
\includegraphics[width=8 cm]{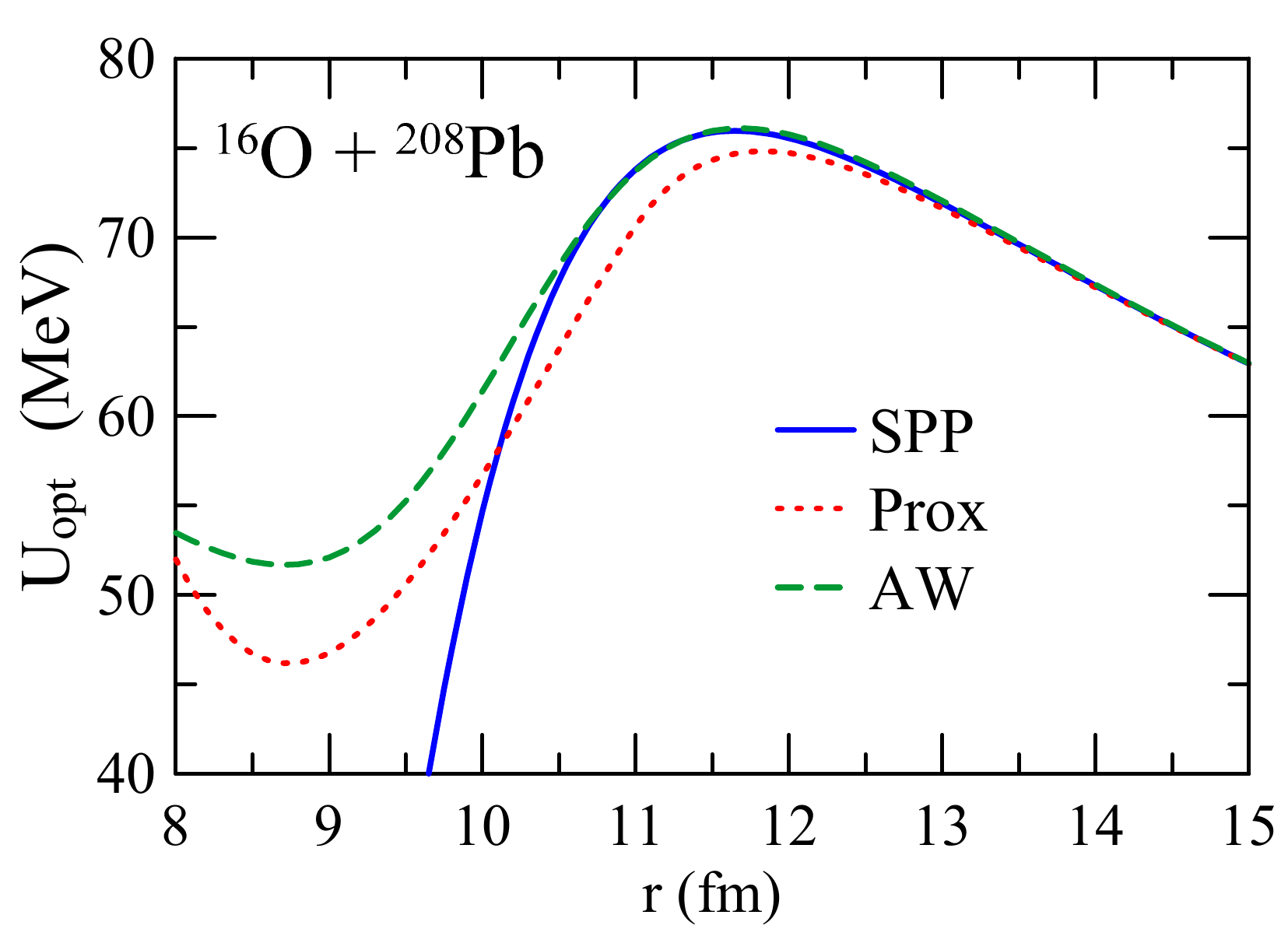}
\caption{(Color online) Coulomb barriers for the bare potentials discussed in the text.}
\label{barriers}
\end{figure}

In this section we look at the Coulomb barriers resulting from the bare potentials of the previous section
and at their influence on the fusion cross sections. As an example, we consider the barriers for the
$^{16}$O + $^{208}$Pb system. The results for the SP, the AW and the Proximity potentials
are are shown in Fig~\ref{barriers}. 

\medskip
\begin{table}
\centering
\begin{tabular} [c] {ccccc}
\hline
& & & &\\ 
systen  & \ \ parameter\   & AW & Prox & SP \\
& & & &\\ 
 \hline
                                      &  $V_{\scr B}$   &  21.1  &   20.0    &  20.8      \\
   $^8$B+$^{58}$Ni       &  $R_{\scr B}$   &   8.9   &   9.3     &   8.9        \\
                                     & $\hbar\omega$ &  4.6   &   4.0      &   4.1        \\
\hline 
                                    &  $V_{\scr B}$    & 20.6   &  {\bf 19.3}     &  {\bf 21.2}  \\
 $^4$He+$^{209}$Bi   &  $R_{\scr B}$    & 10.9   &   {\bf 11.6}    &   {\bf 10.6}  \\
                                    & $\hbar\omega$ &  5.3    &  4.7       &  5.6      \\
\hline       
                                   &  $V_{\scr B}$    &  21.6 &  20.6  & 20.9 \\
$^6$He+$^{238}$U   &   $R_{\scr B}$   & 11.6  &  12.1  & 11.9 \\
                                  & $\hbar\omega$ &  4.5  &   3.8   & 4.0   \\
\hline 

                                 &  $V_{\scr B}$     & 30.4  &  29.0  & 29.8  \\
$^6$Li+$^{209}$Bi   &   $R_{\scr B}$    & 11.1  &  11.6  &  11.3  \\
                                 &  $\hbar\omega$ & 5.2   &   4.6   &   4.8  \\
\hline
                                      &  $V_{\scr B}$     & 83.8  & {\bf 82.6}   &  {\bf 83.9}  \\
$^{32}$S+$^{100}$Mo  &   $R_{\scr B}$    & 10.8  & 10.9   &  10.7  \\
                                      &  $\hbar\omega$ & 3.9    & 4.1     &  4.0    \\
\hline
                                        & $V_{\scr B}$      & {\bf 139.6} & {\bf 138.2} & 139.2 \\
$^{48}$Ca+$^{154}$Sm  &   $R_{\scr B}$    & 12.0   & 12.1   & 12.0  \\
                                        &  $\hbar\omega$ &  3.6    & 4.2      & 3.9 \\   
\hline                                 
\end{tabular}
\caption{
Barrier parameters for different systems. $V_{\scr B}$ and $\hbar\omega$ are given in MeV
and $R_{\scr B}$ is given in fm. We use boldface to indicate parameters which are very different
for different potential models.
}
\label{barpar}
\end{table}
Clearly the Coulomb barriers generated by these potentials are quite different, especially in the inner region. They 
differ by their height and also by their widths. Thus, they are expected to lead to different tunneling probabilities at 
low energies. 
Similar differences can be found for other systems. To illustrate this point, Table~\ref{barpar} shows
the barrier height ($V_{\scr B}$), radius  ($R_{\scr B}$) and curvature ($\hbar\omega$)
parameters for the systems: $^{8}$B+$^{58}$Ni, $^{4}$He+$^{209}$Bi, $^{6}$He+$^{238}$U, $^{6}$Li+$^{209}$Bi, $^{32}$S+$^{100}$Mo and $^{48}$Ca+$^{154}$Sm. 
Clearly there are important differences in parameters associated with each of the bare potentials considered in our
discussion. 

\medskip

In situations where two potentials produce very different barriers, one expects significant differences in the resulting 
fusion cross section. To check this point, we investigate the theoretical cross sections obtained with the above bare
potentials for two systems: $^{6}$He + $^{238}$U and $^{4}$He + $^{209}$Bi. The barrier parameters associated with 
each of these potentials are considerably different, specially for the latter system. 

\begin{figure}[ptb]
\centering
\begin{minipage}[ht] {1\linewidth}
\centering
\subfigure{
\includegraphics[width=7.5 cm]{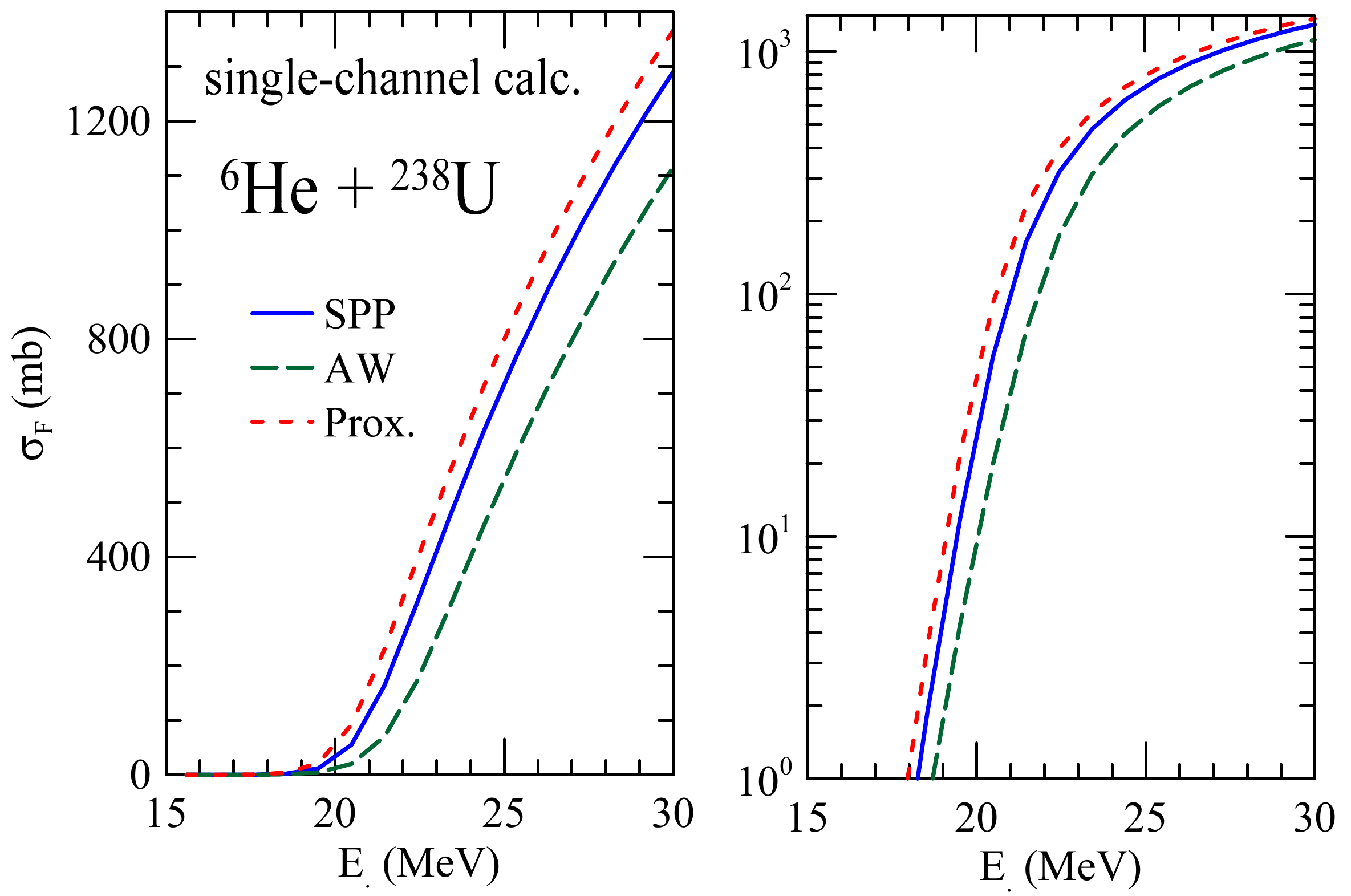}
}
%
\subfigure{
\includegraphics[width=7.5 cm]{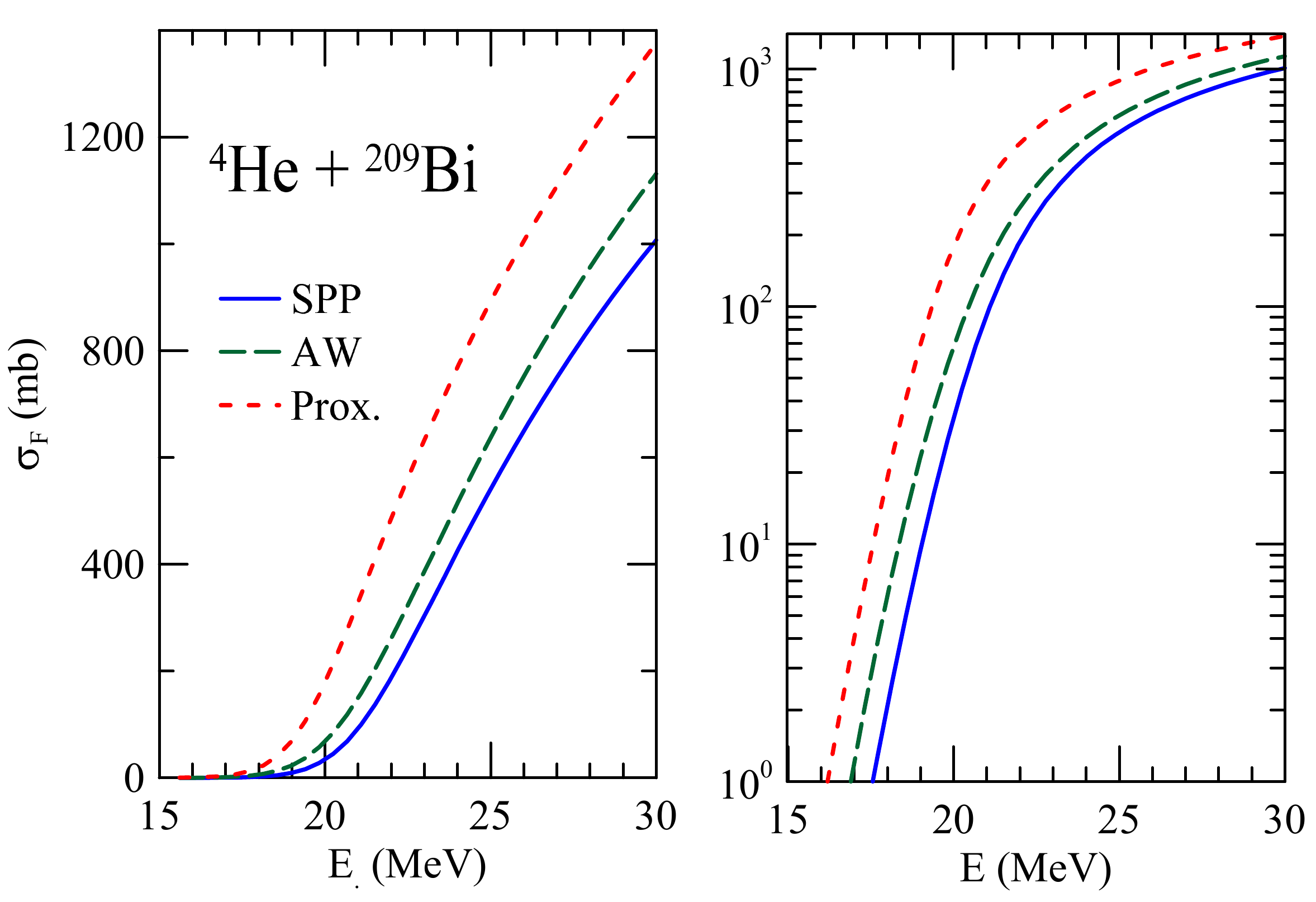}
}
\caption{(Color online) Fusion cross sections for the $^{6}$He + $^{238}$U (top panels) and $^{4}$He + $^{209}$Bi
(botom panels).
The cross sections where obtained by single-channel calculations (no-couplings) using each of the bare potentials
discussed in the text. To see clearly the behaviors above and below the barrier, the results are given in linear (left
panels) and logarithmic (right panels) scales.}
\label{OM_crossec}
\end{minipage}
\end{figure}

The fusion cross section obtained by single-channel calculations (without channel-couplings) using the SP (solid lines),
the AW (long-dashed lines) and the Proximity (short-dashed lines) potentials are shown in Fig.~\ref{OM_crossec}. 
We notice  clearly significant differences among the calculation with the three bare potentials. In particular the $^{4}$He
case, the differences are indeed quite large. The fusion cross sections of the Borromean nucleus $^{6}$He calculated with the SP and
the Proximity potential are close, but that obtained with the AW  is much lower. Obviously in coupled channels analysis of fusion, the 
conclusions concerning the importance of the coupling is intimately related to the bare potential employed. Great care 
must be exercised in reaching these conclusions. What bare potentials one should use, will depend on a concomitant 
analysis of the cross section of other processes which are also sensitive to the barriers, such as breakup and elastic 
scattering. 

\smallskip

\begin{figure}[th]
\centering
\includegraphics[width=8.5 cm]{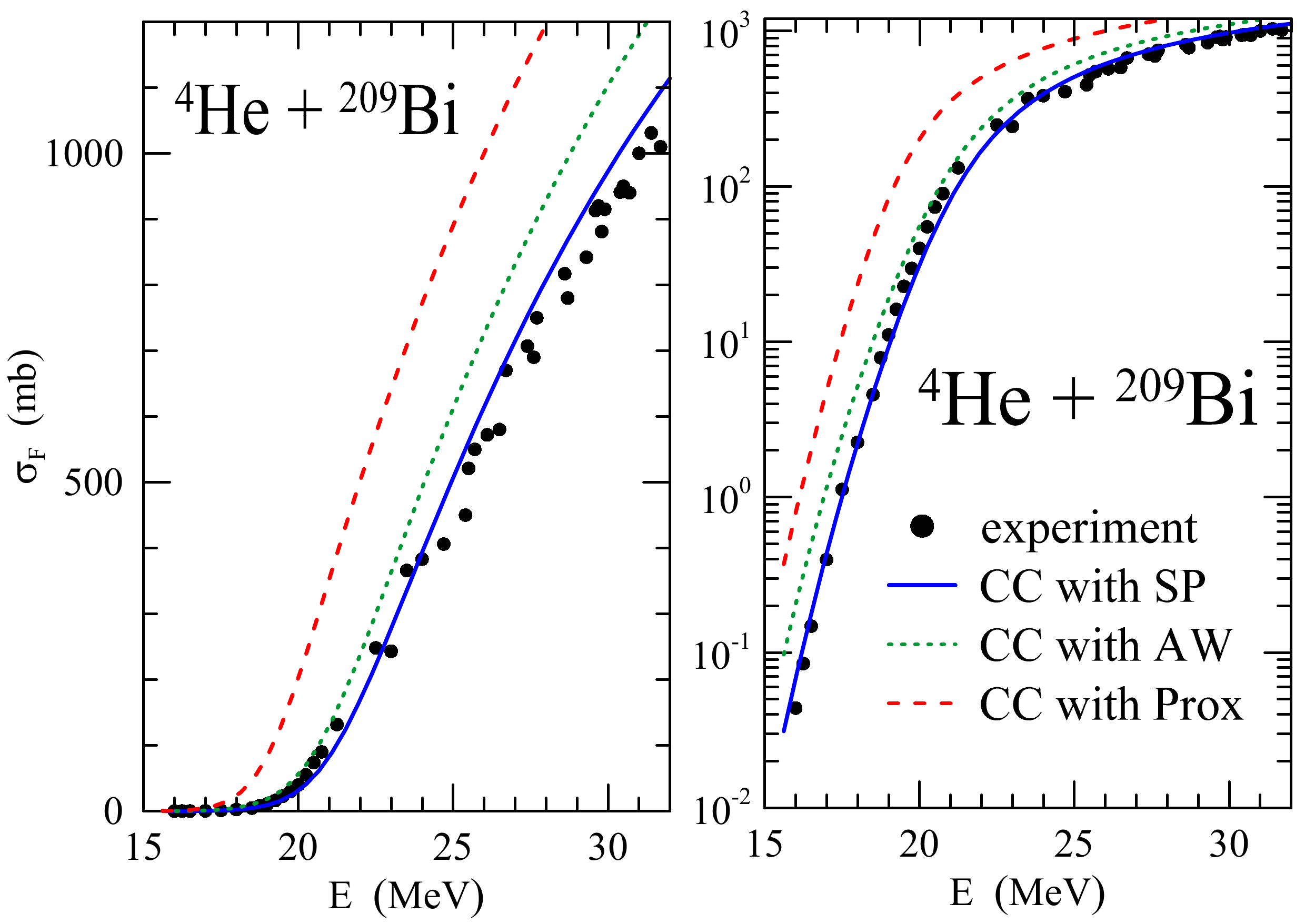}
\caption{(Color online) Comparison of the experiment fusion cross section for the $^{4}$He + $^{209}$Bi system
with theoretical cross sections obtained by coupled channel calculations using each of the bare potentials discussed 
in the text. As in the previous figures, the results are shown in linear (left panel) and logarithmic (right panel) scales.}
\label{ET_crossec}
\end{figure}

A comparison with experimental fusion cross sections could indicate which bare potential gives more realistic
results. However, comparing the above cross sections with data may be misleading, since these calculations do not
incorporate coupled-channel effects, which are very important, specially for collision energies below the Coulomb
barrier. Performing coupled channel calculations for the $^6$He projectile is too complicated as the continuum of
the three-body system ($^4{\rm He}+n+n$) is too hard to take into account. However, reliable coupled channel
calculations for the $^{4}$He+$^{209}$Bi system can easily be performed and the predictions of each bare
potential can be compared with the experiment. We have performed coupled channel calculations including the 
couplings with all relevant channels, associated with the collective excited states of $^{209}$Bi. The results are 
shown in Fig.~\ref{ET_crossec}. The comparison indicates that the SP potential gives the best description of the data. The 
results of the  Ak\"yuz-Winther potential are also quite satisfactory whereas those for the Proximity potential are 
much poorer.

\medskip

To compare the above results with those involving other systems in an appropriate way, one should reduce the
data, in order to eliminate the differences arising from size and charges of the collision partners. There are different proposals to reach this
goal. A convenient one is to transform the cross sections into fusion functions, which depend on
a dimensionless variable associated with the collision energy. One carries out the transformations~\cite{CGL09,CGL09a}:
\begin{equation}
E \, \rightarrow\, x = \frac{E - V_{\scr B}}{\hbar \omega}\ \ {\rm and}\ \ 
\sigma_{\scr F}\, \rightarrow\, F(x) = \frac{2E}{\hbar\omega\, R_{\scr B}^{2}} \times  \sigma_{\scr F}.
\label{fusfun}
\end{equation}
If coupled channel effects can be neglected and Wong's approximation~\cite{Won73} for the cross section holds,
one obtains the Universal Fusion Function (UFF),
\begin{equation}
F_0(x) = \ln{\left[ 1 + \exp{(2\pi x)} \right]} .
\label{UFF}
\end{equation}
For light systems ($Z_{\scr P}Z_{\scr T}\ll 500$) Wong's approximation is poor at sub-barrier energies. However,
this approximation is much better above the barrier and the fusion functions are always close to the UFF in this 
energy region.

\smallskip

It is important to stress that the transformations of Eq.~(\ref{fusfun}) depend on the barrier parameters of
the bare potential. Usually, channel coupling effects have a major influence on the fusion cross section at sub-barrier 
energies. However, frequently, this influence is not so strong above the barrier. Thus, if the bare potential is correctly
chosen, the experimental fusion function, obtained using the experimental cross section in Eq.~(\ref{fusfun}), should
be close to the UFF. That is,
\begin{equation}
\frac{2E}{\hbar\omega\, R_{\scr B}^{2}} \times  \sigma_{\scr F}^{\scr exp}\simeq F_0(x).
\end{equation}
\begin{figure}[th]
\centering
\includegraphics[width=8.5 cm]{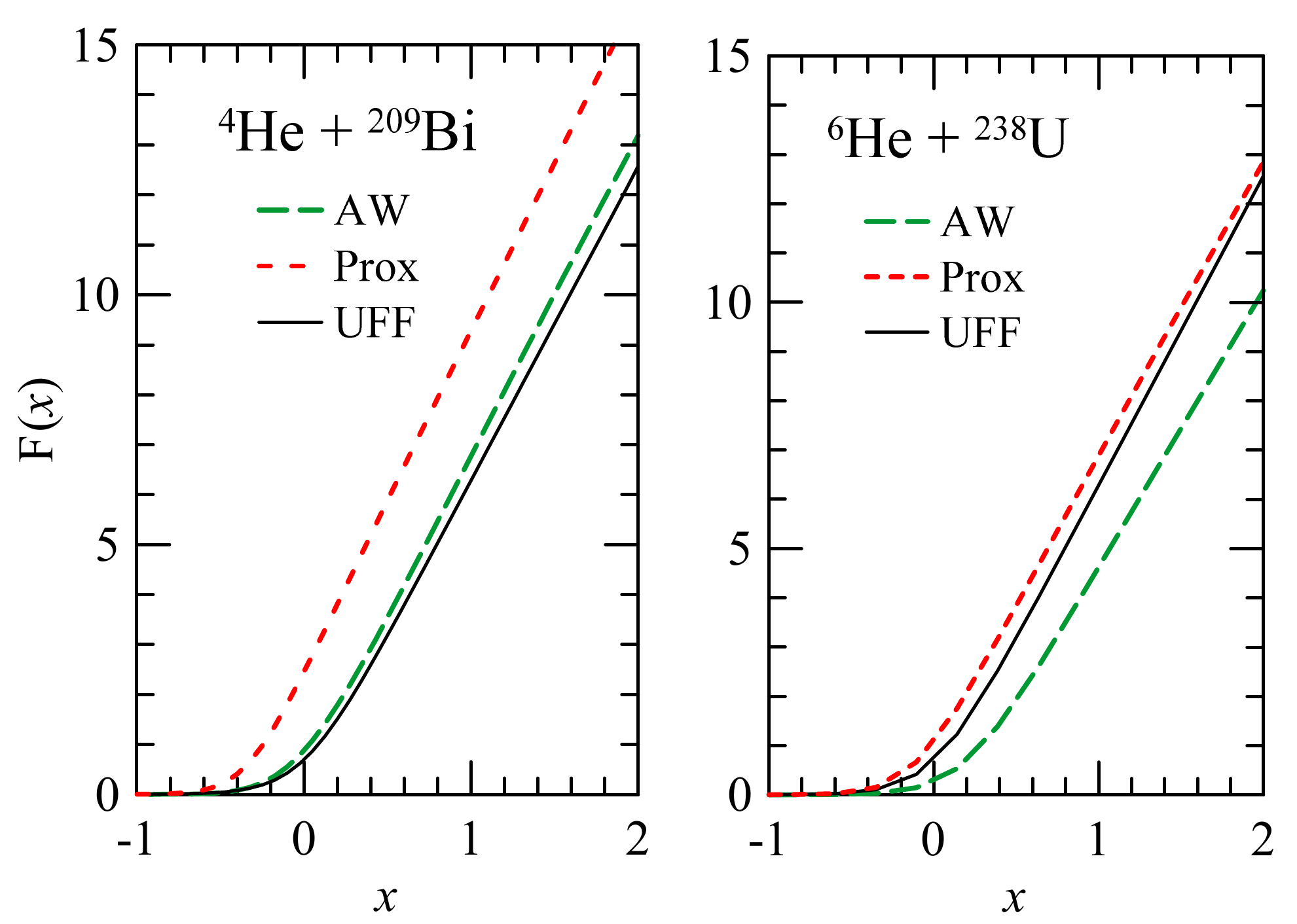}
\caption{(Color online) Fusion functions for the AW and the Proximity potentials, obtained through
single-channel calculations, as explained in the text.}
\label{reduced_OM}
\end{figure}

\bigskip

We use a similar procedure to compare the cross sections of Fig.~\ref{OM_crossec} in terms of fusion functions.
We consider the same systems and use the barrier parameters of the SP potential to carry out the reduction. The fusion 
functions associated with theoretical cross sections obtained by single channel calculations with the AW and 
Proximity potentials are shown in Fig.~\ref{reduced_OM}. We leave out the fusion function associated with the 
SP potential itself, because it is almost indistinguishable from the UFF. This is because Wong's approximation is very
accurate in the above-barrier region shown in detail in the figure. If the AW and the Proximity potentials were 
equivalent to the SP one, their fusion functions would also be very close to the UFF.  We see clearly the sensitivity of 
$F(x)$ to the bare potential employed. In the $^{4}$He case, the calculation of $F(x)$ with the AW bare potential 
is close to the UFF,  while $F(x)$ calculated with the Proximity bare potential it is quite different from UFF. The 
situation is inverted in the $^{6}$He fusion case. Here the AW bare potential gives a $F(x)$ much lower than the 
UFF, while the Proximity potential is practically identical with the UFF. This finding only strengthens our argument 
that there is great sensitivity of near-barrier fusion to the bare potential.  A bad choice of the bare potential may 
lead to wrong conclusions about the behavior of $\sigma_{\scr F}$.

\medskip

\begin{figure}[ptb]
\centering
\begin{minipage}[ht] {1\linewidth}
\centering
\subfigure{
\includegraphics[width=0.78\linewidth]{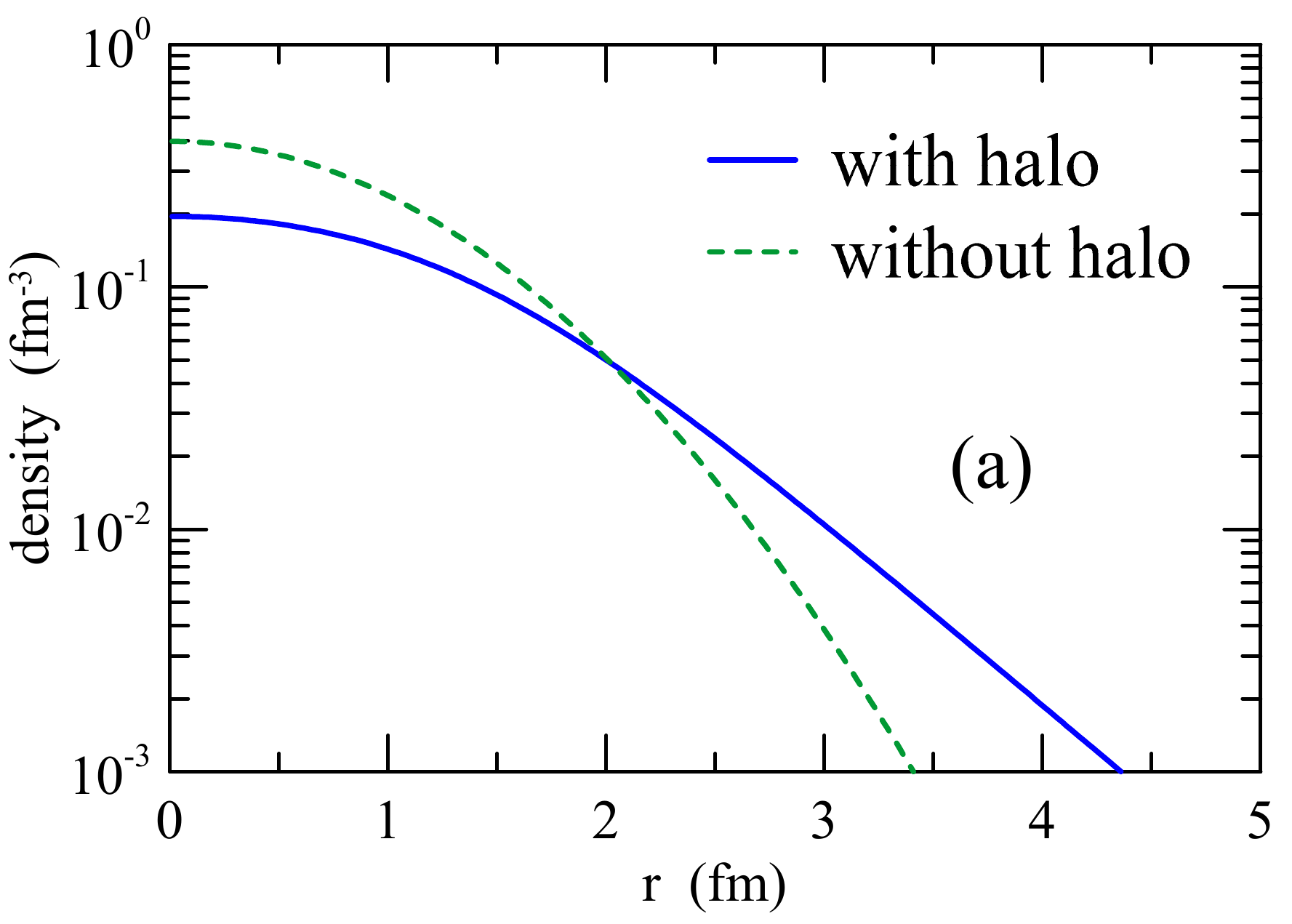}
}
\subfigure{
\includegraphics[width=0.8\linewidth]{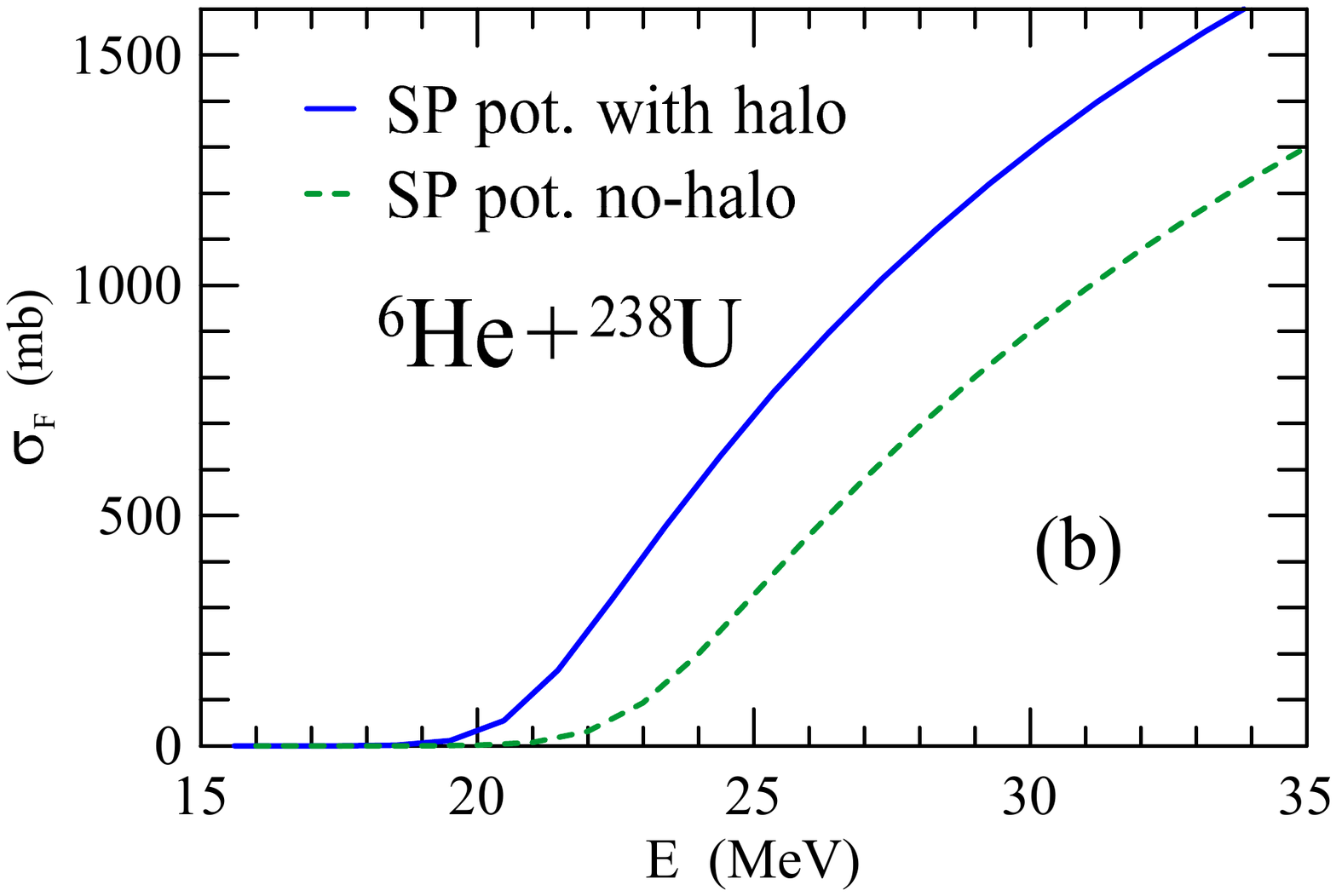}
}
\caption{(Color online) panel (a):  Densities of $^6$He taking into account the two-neutron halo structure (solid line), and treating 
it as a tightly bound nucleus. For details, see the text;  panel (b): Fusion cross sections for the $^6$He+$^{238}$U system obtained from 
single-channel calculations with the SP potential, using the two densities of the previous figure.}
\label{densities-6He}
\end{minipage}
\end{figure}

What causes the differences among the bare potentials used in this paper? We believe the major difference resides in the 
fact that the SP potential, which is the double folding interaction with realistic densities, is more faithful in catching the essentials 
of the nucleus-nucleus interaction. In fact, it is based on experimentally determined densities and there are no approximations 
involved, in contrast to the AW potential, which uses an analytical form of these densities, and employs further approximations to 
allow an analytical evaluation of the double folding integral. The Proximity potential, being constructed for epidermic systems, 
with diffuseness much smaller than the radius of a given nucleus, is reasonable for the heavy targets considered, but fails for 
the light projectile. The systems studied in the previous figures have very unusual densities. On the one hand, $^4$He is a very
compact nucleus whereas the density of $^6$He is influenced by the two neutron halo. To illustrate this point, we compare
the experimental densities of $^6$He~\cite{DJD87}, used in the SP potential, with a gaussian density that does not include the contribution of the
halo. This schematic density is given by a gaussian normalised by the condition $A=6$ and with r.m.s. radius corresponding to that
of $^4$He scaled by the factor $(6/4)^{1/3}$, as in Ref.~\cite{CCC03}. The two densities are compared in Fig.~\ref{densities-6He} (a).
They are indeed very different. The fusion cross sections obtained with the SP potential using these densities are exhibited in 
Fig.~\ref{densities-6He} (b). As expected, the cross sections are considerably different. This clearly shows the importance of using
realistic densities in the calculation of the bare potential.

%





\section{Extracting barrier parameters from above-barrier data}

Attempts have been made to infer the parameters of the bare potential using fusion cross section measurements at 
above-barrier energies (see, e.g. Ref.~ \cite{KGP98}). If one uses the classical approximation for the transmission factor,
\begin{eqnarray}
T_l(E) & = & 1, \qquad {\rm for}\ E > V_{\scr B}(l)\\
           & = & 0, \qquad {\rm for}\ E \le V_{\scr B}(l),
           \label{Tlclass}
\end{eqnarray}
where $V_{\scr B}(l)$ stands for the height of the barrier of the effective potential at the angular momentum $l$, 
the partial-wave series can easily be evaluated. One then gets the fusion cross section,
\begin{equation}
\sigma_{\scr F} = \pi \, R_{\scr B}^2 \ \left[ 1 - \frac{V_{\scr B}}{E} \right],
\label{classical_xsec}
\end{equation}
with $V_{\scr B}\equiv V_{\scr B}(l=0)$. Thus, plotting $\sigma_{\scr F}$ vs. $1/E$, the barrier height and radius can be immediately obtained from the interceptions
of the straight line with the two axis. It intercepts the $1/E$ axis at $1/V_{\scr B}$ and the $\sigma_{\scr F}$ axis at $\pi R_{\scr B}^2$. 

\smallskip

Before further discussions of this procedure, one has to decide how much above the barrier one should take the data. Above some critical 
energy, the highest angular momenta should be excluded from the partial-wave summation. This is because at these angular momenta
the effective optical potential,
\begin{equation}
V_l(r)=V_{\scr N}(r)+V_{\scr C}(r)+\frac{\hbar^2}{2\mu\,r^2}\,l(l+1),
\end{equation}
with $V_{\scr N}(r)$ and $V_{\scr C}(r)$ standing respectively for the bare nuclear potential and the Coulomb interaction, does not have 
a pocket. Thus the projectile cannot be captured and fusion does not take place. 
The fusion cross section is then in another energy regime, where it does not  have a linear 
dependence on $1/E$. On the other hand, at energies just above the barrier, the classical approximation for the transmission factor is 
inaccurate. To choose the energy region where the cross section has the desired linear behavior, one should compare the cross section 
of Eq.~(\ref{classical_xsec}) with its quantum mechanical counterpart. In order to get a system independent answer, we make this
comparison using reduced cross sections. That is, we compare the UFF with its high energy limit, where tunnelling effects can be
neglected. It can be easily checked that at high enough energies ($x  \gg 1$) Eq.~(\ref{UFF}) can be approximated by its limit, 
\begin{equation}
F_0(x\rightarrow \infty) \rightarrow F_0^{\rm cl}(x)= 2\pi\, x.
\end{equation}
\begin{figure}[th]
\centering
\includegraphics[width=6.0 cm]{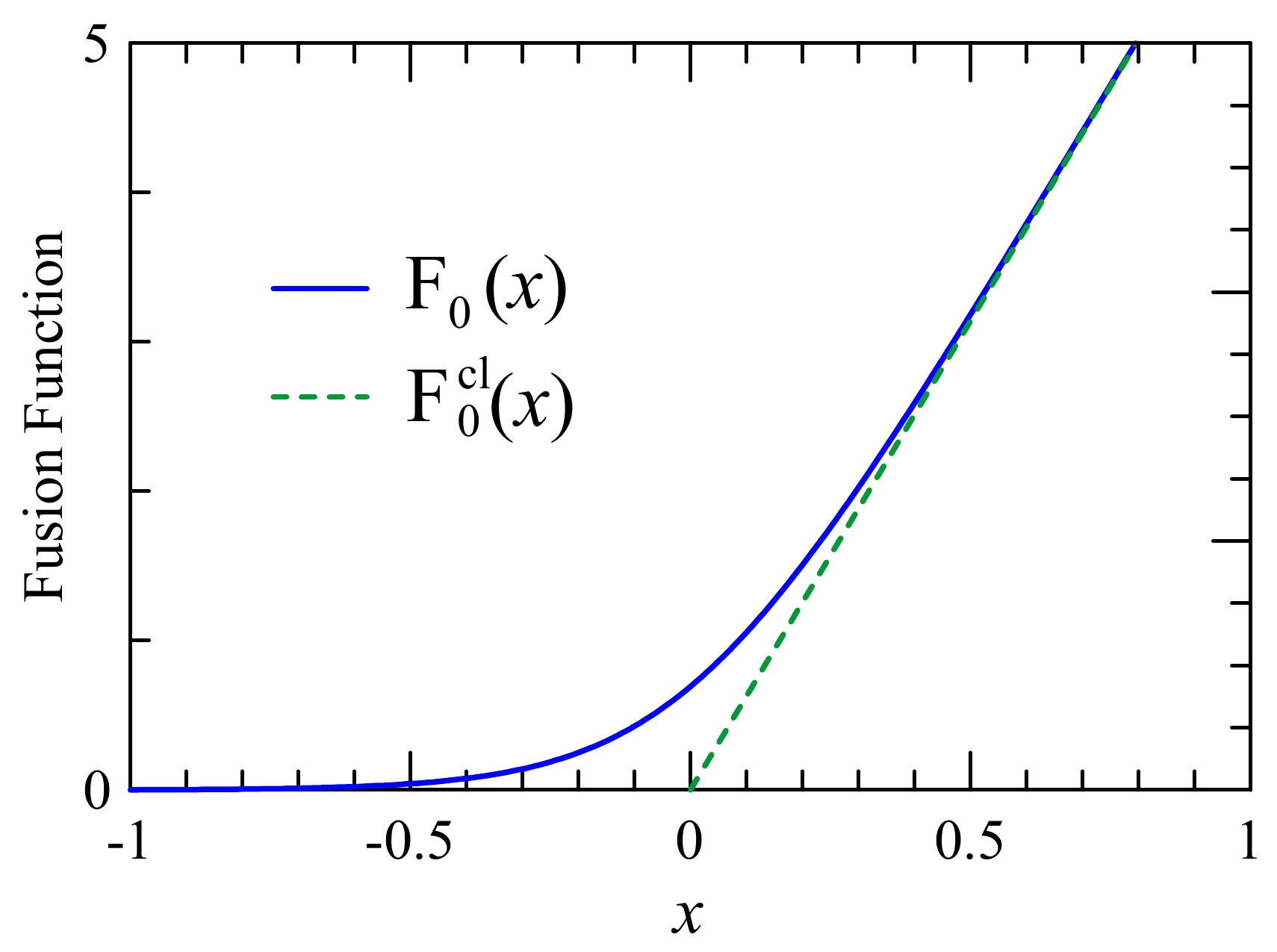}
\caption{(Color online) The Universal Fusion Fusion (solid line) and its high energy limit (dashed line).}
\label{UFFclass}
\end{figure}

In Fig.~(\ref{UFFclass}) we compare the exact form of the UFF with its asymptotic limit. The comparison indicates that the asymptotic
expression can be safely used
for $x > 0.5$. That is, for energies $\hbar\omega/2$ above the barrier. Since $\hbar\omega \sim 4$ MeV, one
should take data starting at about 2 MeV above the barrier.

\medskip

\begin{figure}[th]
\centering
\includegraphics[width=8.0 cm]{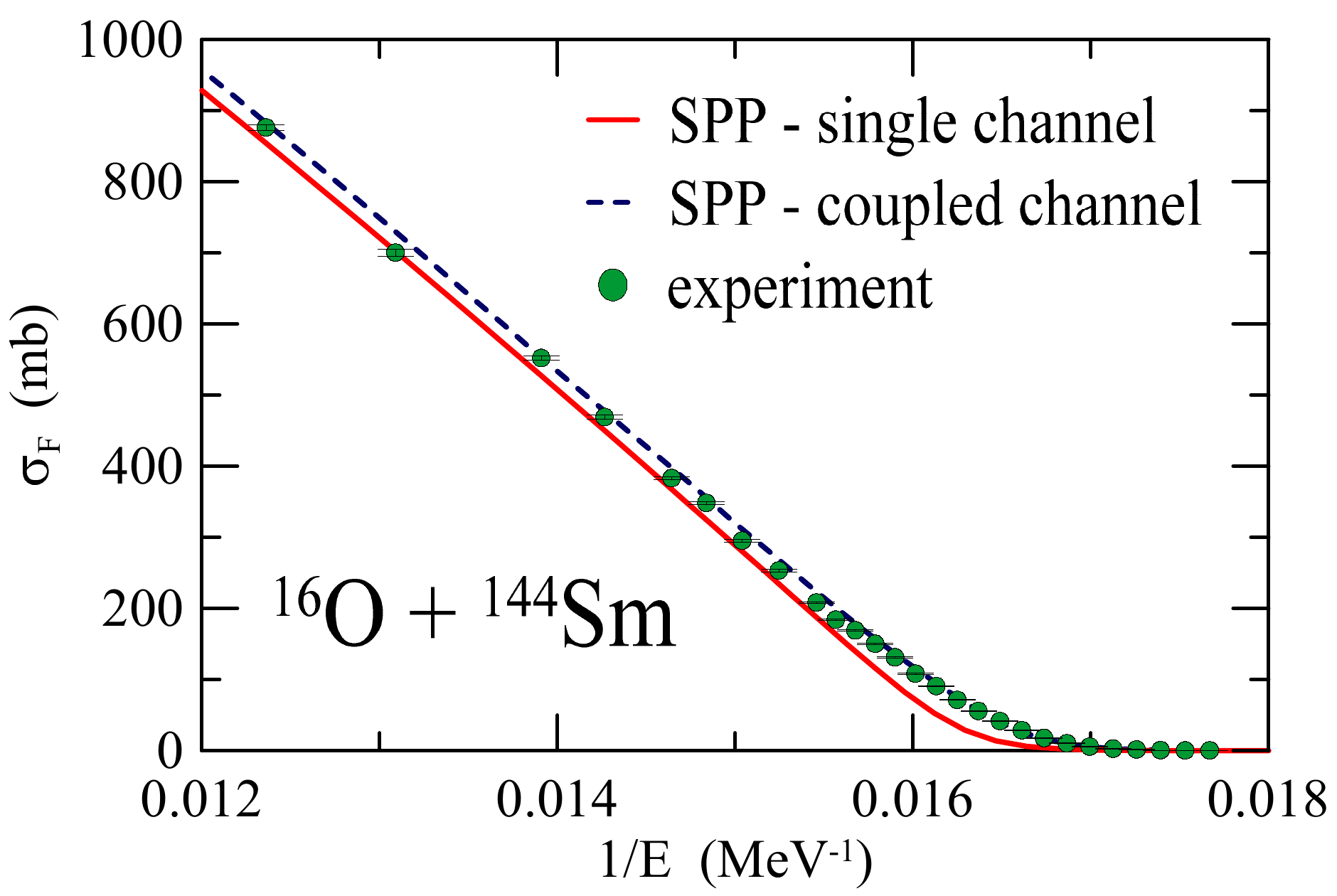}
\caption{(Color online) Fusion cross sections for the  $^{16}$O+$^{144}$Sm system plotted against $1/E$. The solid line corresponds to 
a single-channel calculation with the SP potential whereas the dashed line are results of a coupled channel calculation using the same bare potential.
The experimental data of Leigh {\it et al.}~\cite{LDH95} are also shown.}
\label{OSm_1overE}
\end{figure}
\medskip
\begin{table}
\centering
\begin{tabular} [c] {ccccc}
\hline
      & \ \ \  parab. fit    &  \ \ \  single-channel & \ \ \  CC &   \ \ \  data  \\
 \hline
$V_{\scr B}$ (MeV)       &   61.4  &  61.2    &  60.6     &   60.8      \\
$R_{\scr B}$ (fm)       &  10.9    &   10.6     &  10.6  & 10.6     \\
 \hline
 \hline                                 
\end{tabular}
\caption{
Barrier parameters for the $^{16}$O+$^{144}$Sm system. $V_{\scr B}$ is given in MeV
and $R_{\scr B}$ is given in fm.
}
\label{param2}
\end{table}

The difficulty with this procedure is that the experimental cross section is usually influenced by channel coupling effects. Although
these effects are much more important below the Coulomb barrier, they may affect the cross section at above-barrier energy. In 
this case, the parameter extracted from the cross section would be associated with some sort of effective barrier without any
practical utility. Thus, the influence of channel couplings above the barrier should be checked. 

\smallskip

As a first step, we consider the trivial case of the $^{16}$O+$^{144}$Sm system, where these effects are known to be very weak.
In Fig.~\ref{OSm_1overE} we compare the experimental fusion cross section of Leigh {\it et al.}~\cite{LDH95} with results of the 
single-channel (solid line) and coupled channel (dashed line) calculations. In both cases, we used the SP as the bare potential.
As expected, the single-channel cross section is close to the coupled channel one and to the data. Fitting straight lines to the linear
region of the cross section, we extracted the height and the radius of the barrier, for the two calculations and for the data. The results
are shown in table \ref{param2}, together with correct values of the barrier parameters. The latter was obtained fitting a parabola
to the barrier of the total optical potential (sum of the bare nuclear potential with the Coulomb interaction). The radii obtained
from the $1/E$ dependence of the two calculations and the data are quite close to the correct radius of the barrier. The three are
identical and they differ from the exact value by 0.3 fm. The barrier heights, however, are not so close. The height extracted from
the data is 0.6 MeV below the correct height. We should keep in mind that the purpose of this procedure is to extract the parameters 
of the bare potential from the linear fit of the data and even in this extreme case of weak coupling the barrier height differs from the 
correct value by 0.6 MeV. This example does not encourage the use of the method in situations where the couplings are not so weak.

\medskip

\begin{figure}[th]
\centering
\includegraphics[width=7.0 cm]{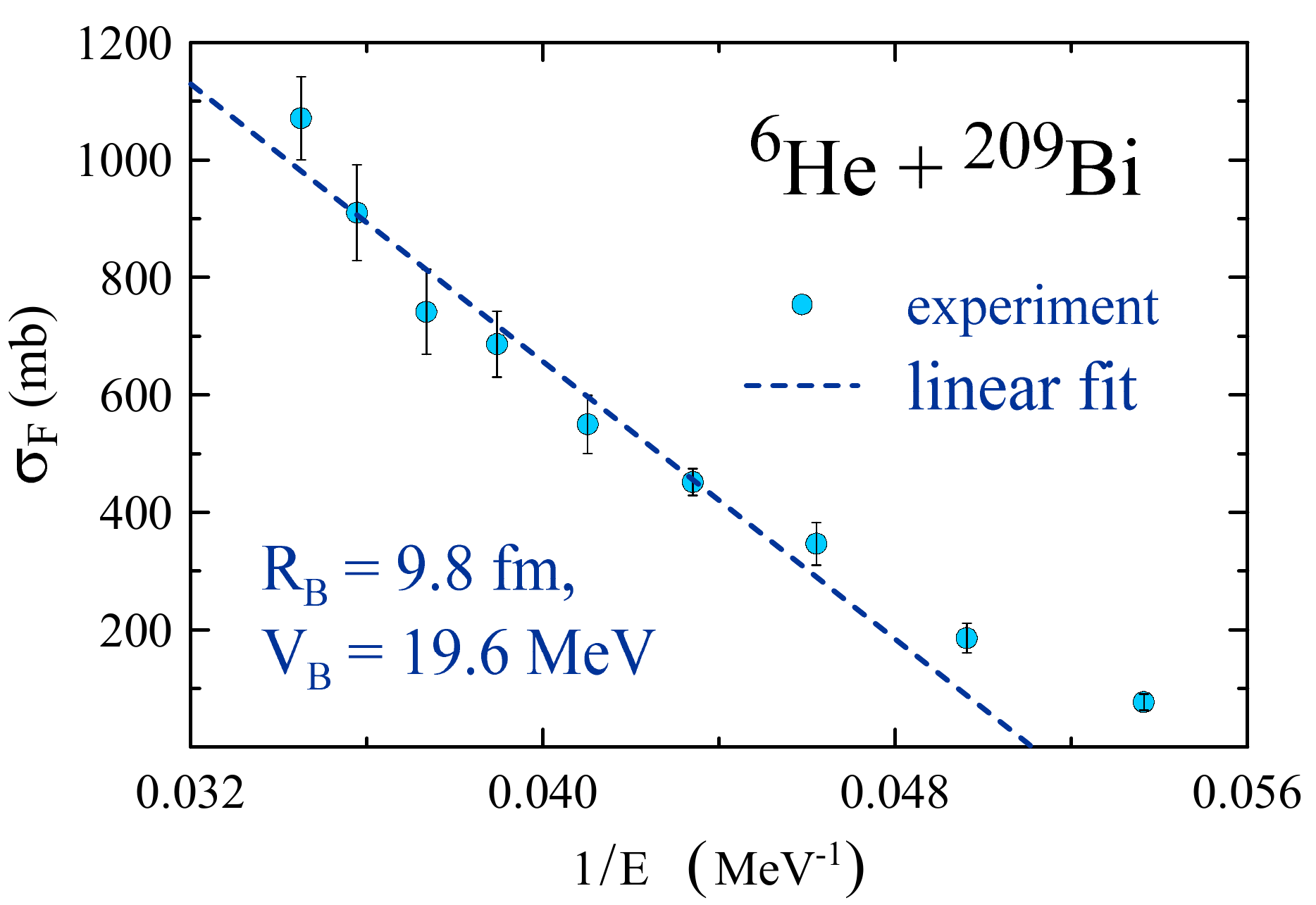}
\caption{(Color online) Experimental fusion cross section for the $^6$He + $^{209}$Bi system~\cite{KGP98}, plotted against $1/E$.
The dashed line corresponds to a linear fit, to the expression of Eq.~(\ref{classical_xsec}).  The barrier height and radius obtained are indicated within the figure.}
\label{HeBi_1overE}
\end{figure}

Now we study another example, where coupling effects are known to influence the cross section above the barrier. We consider
collisions of the halo-nucleus $^6$He with a $^{209}$Bi target. In Fig.~\ref{HeBi_1overE} we show the data of 
Kolata {\it et al.}~\cite{KGP98} plotted against $1/E$, and the corresponding linear fit. From this fit we have obtained the
radius and the height of the barrier. The results are shown in table \ref{param3}, in comparison with the barrier parameters
associated with the three potentials of section II. In each case, the parameters were obtained fitting a parabola to the barrier
of the total optical potential. 
\begin{table}
\centering
\begin{tabular} [c] {ccccc}
\hline
      &  \  \ \ \ \ AW    &  \ \ \  Prox & \ \ \  SP &   \ \ \ experiment  \\
 \hline
$V_{\scr B}$ (MeV)  \qquad     & \qquad  20.0   &  19.0    &  20.9     &   19.6      \\
$R_{\scr B}$ (fm)    \qquad       &  \qquad 11.3    &   11.8     &  11.9  &   9.8     \\
 \hline
 \hline                                 
\end{tabular}
\caption{
Barrier parameters extracted from above-barrier data of the $^6$He + $^{209}$Bi system, in comparison with the corresponding
values for the three bare potentials discussed in the text.
 }
\label{param3}
\end{table}
The discrepancy between the barrier height extracted from the data and that for the SP potential is not larger than the 
discrepancies among the predictions of the three bare potentials.
However, $1/E$ fit leads to a very wrong barrier radius. Independently of the option for the bare potential, the radius
taken from the above-barrier data is $\sim 2$ fm too small. This deviation arises from the strong suppression of the fusion
cross section above the barrier, arising from breakup couplings~\cite{CGL09,CGL09a}.

\smallskip

\section{Conclusions}

In this paper we report on a detailed study of the bare optical potentials frequently employed in the analysis of near-barrier heavy-ion fusion reactions. 

We have found that these potentials give rise to different fusion barriers. They consequently lead to different conclusions about the effect of the channel couplings when coupled channels calculation of the fusion cross section is perforrmed. 
We have further found that folding potentials based on realistic densities tend to be more reliable. 

We have further shown that the extraction of $R_{\scr B}$, $V_{\scr B}$ from $1/E$ plots may be accurate within $\sim 0.5$ fm and $\sim 0.5$ MeV when 
channel coupling effects above the barrier are very small. However, if coupled channels affect the fusion cross section above the barrier, this 
method may lead to meaningless barrier parameters.

\medskip

\textbf{Acknowledgements}

\medskip \noindent The authors would like to thank the financial support
from CNPq, CAPES, FAPERJ, FAPESP and the PRONEX.

\medskip


%
%
\end{document}